\documentclass[aps, reprint, groupedaddress,  superscriptaddress, amsmath, amssymb,onecolumn,prl]{revtex4-2} 
\usepackage{amssymb}
\usepackage{verbatim}
\usepackage{graphicx}
\usepackage{subfigure}
\usepackage{epstopdf}
\usepackage{braket}
\usepackage{amsmath}
\usepackage{soul}
\usepackage{xcolor}
\usepackage{varwidth}
\usepackage{newtxmath}

\usepackage[colorlinks=true,linkcolor=blue,citecolor=blue,urlcolor=blue]{hyperref}

\begin{document}

\title{Critical Role of Hydrogen in Unconventional Superconductors: The Case of Hydrogenated FeSe Layers}

\author{Lan-Lin Du$^\#$}
\affiliation{Beijing National Laboratory for Condensed Matter Physics and Institute of Physics, Chinese Academy of Sciences, Beijing, 100190, P. R. China}
\affiliation{School of Physical Sciences, University of Chinese Academy of Sciences, Beijing 100049, P. R. China}

\author{Yang Yang$^\#$}
\affiliation{Beijing National Laboratory for Condensed Matter Physics and Institute of Physics, Chinese Academy of Sciences, Beijing, 100190, P. R. China}
\affiliation{School of Physical Sciences, University of Chinese Academy of Sciences, Beijing 100049, P. R. China}

\author{Shiqi Hu}
\affiliation{Beijing National Laboratory for Condensed Matter Physics and Institute of Physics, Chinese Academy of Sciences, Beijing, 100190, P. R. China}

\author{Sheng Meng}
\email{smeng@iphy.ac.cn}
\affiliation{Beijing National Laboratory for Condensed Matter Physics and Institute of Physics, Chinese Academy of Sciences, Beijing, 100190, P. R. China}
\affiliation{School of Physical Sciences, University of Chinese Academy of Sciences, Beijing 100049, P. R. China}
\affiliation{Songshan Lake Materials Laboratory, Dongguan, Guangdong 523808, P. R. China}

\begin{abstract}
Hydrogenation is known to tune superconductivity in a wide range of materials. While its microscopic role has been clarified in phonon-mediated superconductors such as hydrogenated MgB$_2$, LaH$_{10}$, and H$_3$S, much less is known for hydrogenated cuprates and iron-based superconductors, where even the underlying structural motifs remain elusive. Using hydrogenated FeSe as a prototypical example, we reveal how hydrogen affects superconductivity in the presence of strong electronic correlations: correlation-induced orbital renormalization shifts hydrogen-derived spectral weight from the high-energy region toward the Fermi surface (FS), 
remarkably enhancing the electron–phonon coupling (EPC). We predict a structurally stable FeSeH phase where, compared to bare FeSe, hydrogen incorporation reshapes the FS topology and increases the number of channels for electron–phonon scattering, while simultaneously introducing high-frequency phonons that strengthen pairing. First-principles EPC calculations combined with dynamic mean field theory (DMFT) yield a superconducting transition temperature ($T_c$) exceeding 40 K. Fully anisotropic Eliashberg theory reveals a two-gap superconducting state, consistent with the gap structure experimentally observed in doped FeSe. Our findings identify correlation-enhanced EPC as a plausible microscopic mechanism for iron-based superconductivity and offer a new perspective on pairing in strongly correlated systems. In addition, this work establishes hydrogenated FeSe as a promising platform for engineering two-dimensional superconductors and superconducting quantum devices.
\end{abstract}

\maketitle

The discovery of new hydrogen-rich 
\cite{Bondarenko2024,Wang2017,Shimizu2020,Pickard2020,Struzhkin2015,Syed2016,Boebinger2024} and two-dimensional (2D) superconductors \cite{Li2021,Qi2011,Qin2023,Qiu2021,Saito2016,Wang2023} is of fundamental importance to both physics and materials science. Hydrogen-rich superconductors, especially under high-pressure conditions \cite{Wang2017,Pickard2020,Struzhkin2015}, are considered crucial for the search of room-temperature superconductivity. The high-frequency phonon modes and strong EPC of hydrogen atoms provide a solid foundation for achieving high superconducting transition temperature $T_c$ \cite{Drozdov2015,Somayazulu2019,Drozdov2019,Peng2017}. For phonon-mediated superconductors such as hydrogenated MgB$_2$ \cite{Bekaert2019}, LaH$_{10}$ \cite{Errea2020} and H$_3$S \cite{Errea2016}, theoretical studies have begun to elucidate the microscopic role of hydrogen in enhancing superconductivity. However, in hydrogenated cuprates \cite{Bobylev2017,Reilly1987} and iron-based superconductors \cite{Cui2019,Cui2018,Xue2024}—representative families of unconventional superconductors—relevant studies are still scarce, and even the essential atomic structural information remains largely undetermined. Meanwhile, 2D superconductors, due to their reduced dimensionality, exhibit unique quantum phenomena absent in conventional three-dimensional superconductors, such as quantum phase transitions \cite{Wang2023} and topological superconducting states \cite{Qin2023,Qi2011}. The intersection of hydrogen-doping and 2D superconductors offers transformative opportunities for both fundamental research of quantum mechanical phenomena and technological innovation in the development of next-generation quantum devices.

Monolayer FeSe has attracted massive attention due to its novel physical properties compared to bulk FeSe \cite{Wang2012,Peng2014,Liu2012,Xue2024}. When epitaxially grown on insulating substrates such as $\rm SrTiO_3$ (STO), monolayer FeSe exhibits a superconducting $T_c$ exceeding 65 K, which is a remarkable enhancement over the $T_c$ of 8~K in bulk FeSe \cite{Hsu2008}. Compared with FeSe/STO, freestanding monolayer FeSe 
circumvents substrate-induced charge transfer, strain, and polar-phonon coupling, allowing direct access to the intrinsic electronic structure and pairing mechanism of 2D FeSe. It also provides a clean platform for external-field tuning and integration into two-dimensional quantum devices. However, achieving pure free-standing monolayer FeSe remains experimentally challenging. The absence of a substrate leads to structural degradation and reconstruction of the FeSe monolayer. Hydrogenation represents a promising and versatile strategy for stabilizing monolayer materials. It involves the addition of hydrogen atoms to the surface or edges of 2D materials, which can passivate reactive sites, reduce surface energy, and prevent undesirable chemical reactions \cite{Moaied2018,Qiu2015,HB2021}. Beyond structural stabilization, emerging experiments suggest that hydrogen can also serve as an effective tuning knob for superconductivity in FeSe-based systems. Protonated bulk FeSe is reported with a superconducting $T_c$ over 40 K \cite{Cui2019,Cui2018}, and hydrogen exposure has been found to enhance superconductivity in monolayer FeSe/STO \cite{Xue2024}. These observations highlight the promise of hydrogenation in FeSe, while the microscopic structure and underlying superconducting mechanism of hydrogenated FeSe remain elusive.

In this work, we use hydrogenated FeSe as a prototypical platform to elucidate how hydrogen promotes superconductivity in a strongly correlated unconventional superconductor. We identify a structurally stable hydrogenated FeSe monolayer and show that hydrogen incorporation enhances superconductivity through intertwined electronic and lattice effects. First, charge transfer from hydrogen to the FeSe layer raises the Fermi level, reconstructs the FS topology, and introduces additional channels for electron-phonon scattering. Second, hydrogen introduces high-frequency phonon modes that can substantially contribute to pairing once electronic correlations are properly included. We employ DMFT, a well-established method for treating strong electronic correlations, to correct the EPC strength calculated from density functional theory (DFT), yielding a superconducting $T_c$ greater than 40 K. The EPC associated with hydrogen-derived phonon modes is strongly enhanced at the DMFT level, compared with standard DFT calculations. Most importantly, our calculations reveal that many-body correlations renormalize the hydrogen $s$-orbital spectral weight from high-energy regions toward the FS, thereby activating strong coupling between low-energy quasiparticles and hydrogen-derived phonons. This finding provides insight into the role of hydrogen in strongly correlated unconventional superconductors, revealing its ability to impact superconductivity through correlation-driven modifications of EPC. Using fully anisotropic Eliashberg theory, we identify that the FeSeH monolayer exhibits two-gap superconductivity, similar to that reported in bulk FeSe \cite{Ponomarev2011,Chen2017,Muratov2018}. The agreement of the calculated two-gap structure and $T_c$ with experimental observations in FeSe-type materials provides valuable insight into correlation-promoted EPC as one of the origins responsible for unconventional superconductivity.

\section{Results}
\subsection{Crystal structure and electronic reconstruction of FeSeH}

To find the stable structure of hydrogenated FeSe, we relax the atomic structures with different number of hydrogen atoms located at different vacancies of both bulk and monolayer FeSe, and calculate the corresponding phonon spectra. Among all of the different atomic configurations, we identify a monolayer one with a 1:1:1 stoichiometry of Fe, Se, and H [FIG. 1(b)], which is dynamically stable as evidenced by the absence of imaginary phonon frequencies in FIG. 1(c) (unstable configurations are presented in FIG. S1 of supplementary materials). In the unit cell, each selenium atom is paired with a hydrogen atom along the out-of-plane direction.

\begin{figure*}
    \centering
    \includegraphics[width=0.8\textwidth]{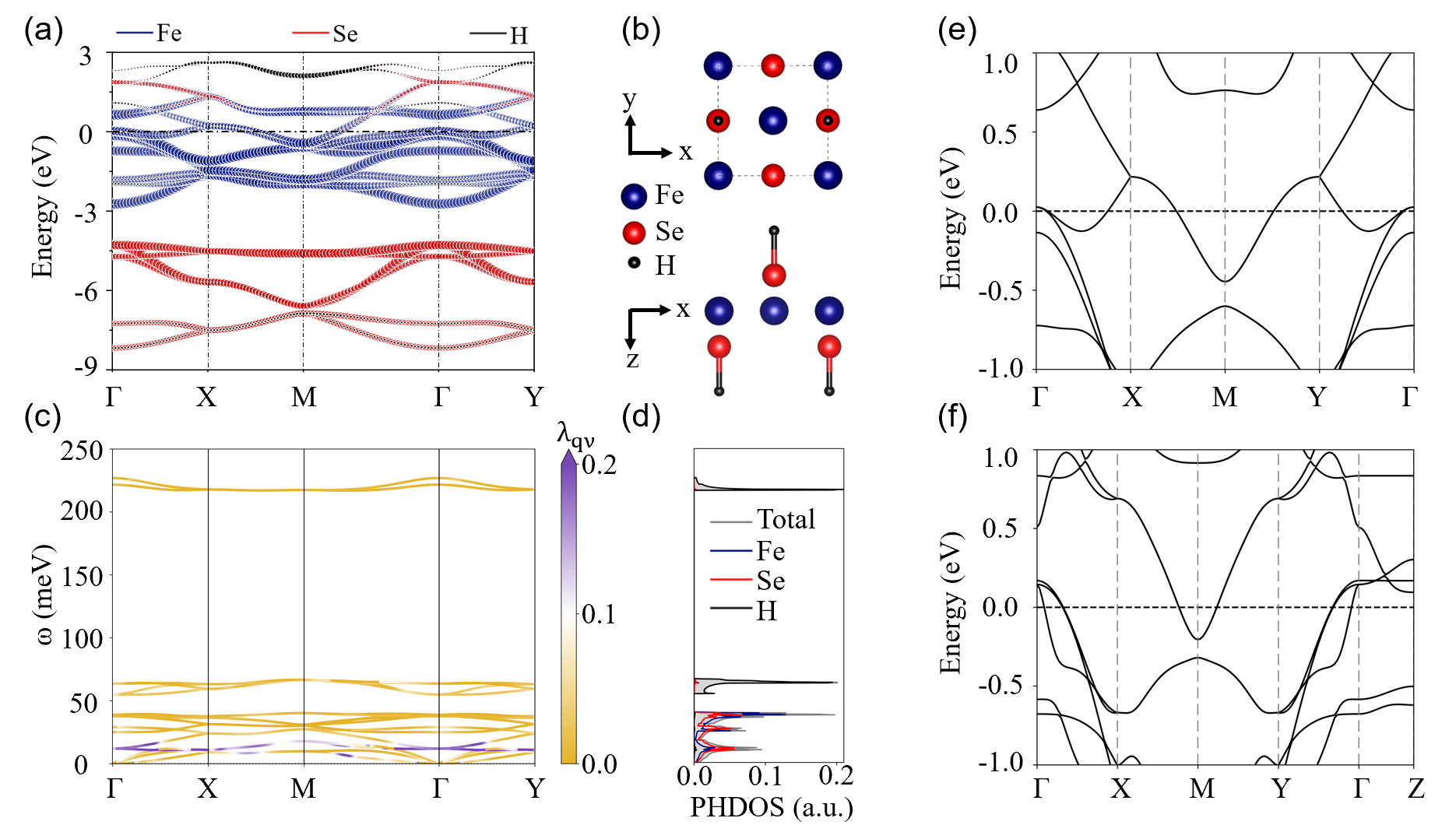} 
    \caption{(a) Fatbands near the Fermi level from DFT, composed of Fe $d$ orbitals (blue), Se $p$ orbitals (red), and H $s$ orbitals (black). (b) Atomic configuration of FeSeH. (c) Phonon dispersions of FeSeH, with the EPC strength ($\lambda_{q,\nu}$) (represented in colorbar). (d) Total and atom-resolved phonon density of states.  Band structure from DFT of (e) FeSeH and (f) bulk FeSe.}
\end{figure*}

FIG. 1(a) shows the fatband near the Fermi level, with 18 electronic bands displayed, composed of Fe $d$ orbitals, Se $p$ orbitals, and H $s$ orbitals. Therefore, an 18-band s-p-d model is required to describe the system, as used in our DMFT calculations. The hydrogen component is located relatively distantly above the Fermi level. The phonon dispersion and density of states are shown in FIG. 1(c) and 1(d), respectively. Similar to other hydrogen-rich compounds, hydrogen atoms contribute to high-frequency phonons. The doping effect of hydrogen atoms causes the Fermi level to shift upward compared to FeSe, as shown in FIG. 1(e) and 1(f), which moves the electron pocket near the M point away from the Fermi level and alters the size and orbital components of the hole pocket near the $\Gamma$ point. In addition, a new electron pocket appears along the $\Gamma$ to X (Y) direction, and a new hole pocket appears near X (Y), which are unambiguously indicated in the Fermi surface of FeSeH [FIG. 2(a)].

\begin{figure}[h] 
    \centering
    \includegraphics[width=0.8\textwidth]{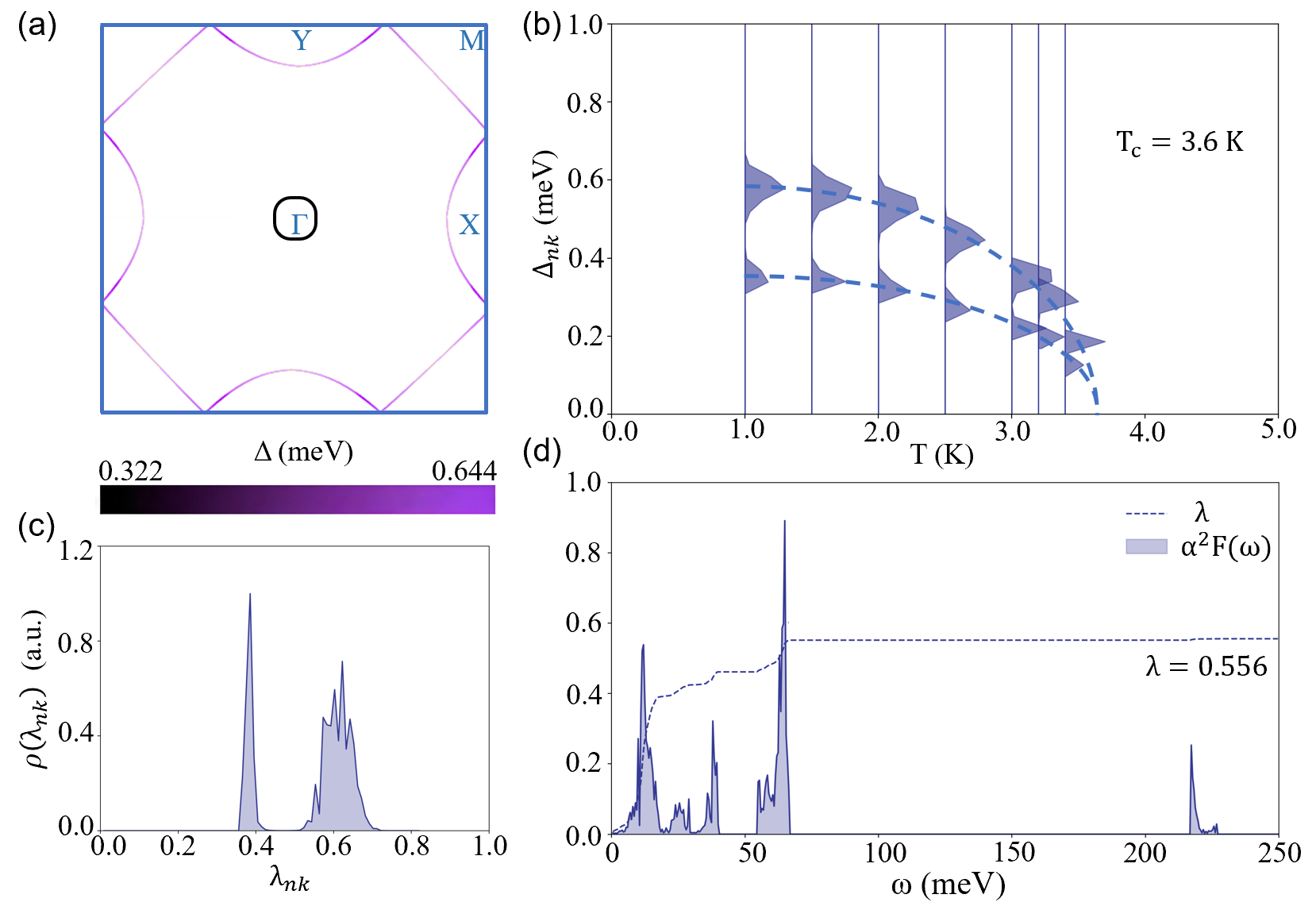} 
    \caption{(a) FeSeH's superconducting gap $\Delta(k)$ as a function of electron momentum $k$  on the Fermi surface at 1~K. The blue square represents the first Brillouin zone. (b) The temperature dependence of the two superconducting gaps. (c) Distribution of the state-resolved EPC strength $\lambda_{nk}$. (d) The frequncy dependence of the isotropically averaged Eliashberg spectral function $\alpha^2F$ and the EPC strength $\lambda$. }
\end{figure}

\subsection{Two-gap superconductivity in FeSeH}
Using density functional perturbation theory (DFPT) and fully anisotropic Eliashberg formula \cite{Savrasov1992,Giannozzi2009,Giustino2007,Giustino2017,Lee2023}, we determine the EPC and superconducting properties of FeSeH. The mode- and momentum-resolved EPC strength $\lambda_{\mathbf{q}{\nu}}$ is shown in FIG. 1(c). The main contributions to the EPC strength, in addition to the $q = 0$ ($\Gamma$  point) corresponding to typical forward scattering in FeSe \cite{Lee2014,Gerber2017,Rebec2017,Rademaker2016,Ding2022}, also arises near the X (Y) points and along the path from the M point to other high-symmetry points. This is directly related to the change in the FS topology discussed above. Namely, the new scattering channels introduced by the newly emerged electron and hole pockets near the Fermi surface  contribute crucially to the EPC strength of FeSeH.

FIG. 2(a) shows the distribution of the
superconducting gap on the Fermi surface at $1$ K. The superconducting gap originates predominantly from the Fe-$d$ orbitals near the FS and the gap values exhibit colossal distinctions around the $\Gamma$ point and near the Brillouin zone boundary, demonstrating the two-gap nature of the superconductivity in FeSeH. We emphasize the consistency between the calculated two-gap feature and existing experimental data of FeSe \cite{Ponomarev2011,Chen2017,Muratov2018}. It suggests that the theoretical framework used to describe phonon-mediated superconductivity captures key features of FeSe-based superconductors. This supports the hypothesis that the superconductivity in FeSe-based materials may arise from electron-phonon coupling, as discussed in a number of recent studies \cite{Fan2015,Lee2014,Tang2016,Zhang2019,Rebec2017,Song2019}.

\begin{figure}[h] 
    \centering
    \includegraphics[width=0.6\textwidth]{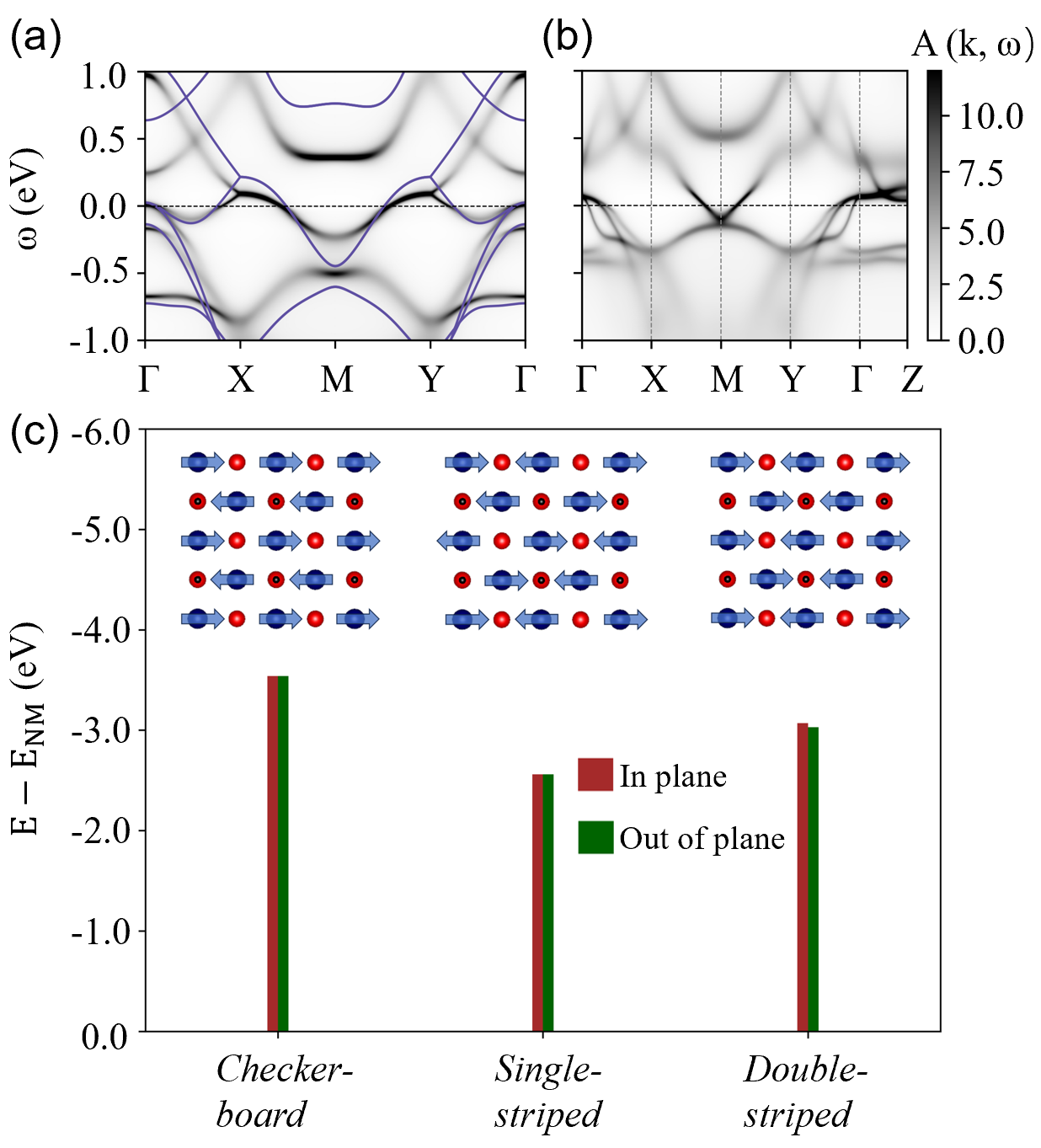} 
    \caption{Spectral function from DMFT of (a) FeSeH and (b) bulk FeSe. The gray shaded line is the spectral function of DMFT, and the purple line is the energy band of DFT. (c) Energies of three antiferromagnetic configurations (checkerboard, single-striped and double-striped) of FeSeH with magnetic moments pointing in plane and out of plane. The atomic configuration and arrows in the upper area indicate the direction of the magnetic moments in the in-plane case. $E_{\mathrm{NM}}$ represents the energy of nonmagnetic configuration.}
\end{figure}

In FIG. 2(d), we display the Eliashberg spectral function $\alpha^{2}F(\omega)$ and the integrated EPC strength $\lambda(\omega)$ of FeSeH. The total EPC strength $\lambda$ is found to be 0.556, significantly higher than the value of 0.15 obtained in bulk FeSe. Our calculated $\lambda$ for bulk FeSe is consistent with previous studies \cite{Zheng2018,Subedi2008}, yielding a negligibly small superconducting $T_c \approx 0~\mathrm{K}$. The two-gap feature is also explicitly demonstrated in both the distributions of the state-resolved EPC strength $\lambda_{nk}$ [FIG. 2(c)] and superconducting gap $\Delta_{nk}$ at different temperatures [FIG. 2(b)]. The zero-temperature amplitudes of these two gaps are approximately 0.36 meV and 0.58 meV, respectively. The temperature at which the superconducting gaps just become zero yields a superconducting $T_c$ of 3.6 K. Although the $T_c$ obtained here is much lower than the experimental values reported for protonated FeSe \cite{Cui2019,Cui2018} and bulk FeSe \cite{Hsu2008}, it is for the first time that DFT gives a non-zero $T_c$ and unambiguously presents two-gap superconductivity in FeSe-based materials. As shown later in our DMFT analysis, electronic correlation effects remarkably enhance EPC and give rise to a $T_c$ comparable to that of protonated FeSe \cite{Cui2019,Cui2018}.

Here we summarize the specific influence of hydrogen on the EPC and superconductivity within the DFT+DFPT framework for FeSeH. First, the DFT fatband analysis shows that the hydrogen s orbital lies around 2 eV above the FS, indicating that hydrogen contribute little to the electron–phonon scattering processes near the FS as shown in FIG. 1(c) and FIG. 2(d) (a picture that will change once electronic correlation is properly included, as discussed later). Instead, its primary effect at the DFT level is to electron-dope the FeSe layer, thereby modifying the FS topology, increasing the number of available scattering channels, and enhancing the electronic density of states near the FS. Moreover, the high-frequency phonons dominated by H atoms further play a beneficial role by increasing the characteristic phonon energy scale relevant for superconducting pairing.

\subsection{Correlation-enhanced electron–phonon coupling}
To explore the modulation of the electron correlation effect on the EPC strength in FeSeH, we used DMFT to calculate its correlation-corrected EPC strength and superconducting $T_c$, following the same theoretical framework as adopted in bulk FeSe \cite{Du2025}. The spectral functions of FeSeH and bulk FeSe are shown in FIG. 3(a) and 3(b), respectively. Compared with the band structure calculated by DFT, the bandwidth of the energy band near the Fermi surface calculated by DMFT is significantly narrower, indicating the electronic correlation effect in FeSeH. The changes of the FS topology in the DMFT spectral function of FeSeH, compared to bulk FeSe, are similar to the DFT results discussed earlier as shown in FIG. 1(e) and 1(f). One notable difference is that the gap between the electron pocket at the M point and the energy band below it remains open in FeSeH, whereas it is closed in bulk FeSe, which reflects the effectiveness of our calculations as discussed in the recent work \cite{Du2025}.

\begin{table*}[!ht]
\caption{Deformation potentials averaged over the Fermi surface within the DMFT ($D_{\nu}^{\text{DMFT}}$, in meV) and DFT ($D_{\nu}^{\text{DFT}}$, in meV), and mode-resolved EPC strength within the DMFT ($\lambda_{\nu}^{\text{DMFT}}$) and DFT ($\lambda_{\nu}^{\text{DFT}}$) for the 15 optical phonon modes.}
\centering
\begin{tabular}{lcccccccccc}
\hline\hline
mode index & 4-5 & 6 & 7& 8-9 & 10-11 & 12 &13-14 &15-16 &17 &18 \\ \hline
$D_{\nu}^{\text{DMFT}}$ & 10.1 & 18.6 & 21.7 & 55.2 & 28.1 & 46.0 & 8.9 & 30.7 & 19.5 & 5.3\\ 
$D_{\nu}^{\text{DFT}}$ & 3.7 & 7.3 & 13.4 & 0.5 & 2.6 & 10.3 & 0.8 & 0.1 & 0.2 & 18.9\\ 
$\lambda_{\nu}^{\text{DMFT}}$ & 1.20 & 0.40 & 0.20 & 2.68 & 0.77 & 0.02 & 2.71 & 4.72 & 0.97 & 0.01 \\ 
$\lambda_{\nu}^{\text{DFT}}$ & 0.33 & 0.04 & 0.05 & 0.0015 & 0.02 & $\sim 0$ & 0.04 &$\sim 0$ & $\sim 0$ & 0.02 \\ 
\hline\hline
\end{tabular}
\end{table*}

We then calculate the deformation potentials of the 15 optical phonon modes at $\Gamma$ point by stretching the atomic positions along the direction of the phonon eigenvectors and calculating the energy difference of the electrons before and after the deformation. These deformation potentials correspond to the EPC matrix elements at $\Gamma$ point, allowing us to determine the mode-resolved EPC strength and, subsequently, the superconducting $T_c$ [see Ref. \cite{Du2025} for more details]. 

As listed in Table I, the deformation potentials calculated by DMFT ($D_{\nu}^{\text{DMFT}}$) are significantly enhanced compared to those calculated by DFT ($D_{\nu}^{\text{DFT}}$). In particular, the deformation potentials of the 8th-11th and 13th-17th phonon modes are enhanced by one or even two orders of magnitude. Examining the eigenvectors of these phonon modes, we find that in the former eight modes, the two H atoms exhibit strong motion in the $xy$ plane, while the 17th mode corresponds to the directional co-moving of the two H atoms along the $z$ direction.

The last two rows of Table I list the mode-resolved EPC strength within DMFT and DFT. The EPC strength calculated by DFT ($\lambda^{\text{DFT}}$) is primarily concentrated in the low-frequency phonon modes, which are mainly associated with the motion of Fe and Se atoms (see FIG. 1(d)). High-frequency phonons, dominated by the motion of H atoms, contribute little to the EPC as discussed above. However, the electron correlation effect included in DMFT greatly promotes the contribution of H vibration to the EPC. 
The final total EPC strength $\lambda^{\text{DMFT}}$ is 2.23, much larger than the value obtained from DFT. Using the EPC strength from DMFT, we calculate a superconducting $T_c$ of 43.7 K, which is very close to the experimentally reported value for protonated FeSe \cite{Cui2019,Cui2018}.

To clarify the correlation-driven enhancement of the EPC for phonon modes dominated by hydrogen motions, we investigate the orbital composition of states near the FS. While H-derived orbitals contribute negligibly to the Fermi-surface states in DFT [FIG. 1(a) and FIG. S2], they acquire significant spectral weight near the FS once many-body correlations are included within DMFT. As shown in FIG. 4, the H orbital components are strongly enhanced, most notably at the hole pockets around the X (Y) point. Consequently, optical phonon modes with large-amplitude hydrogen motions generate sizable deformation potentials associated with these hole pockets (FIG. S3). This correlation-induced renormalization of the orbital character near the FS underlies the pronounced enhancement of EPC and the high superconducting $T_c$ in FeSeH. As shown in FIG. 4, the H‘s spectral weight does not form an independent band but strictly overlaps with the Fe's dispersion. This indicates that the emergent weight near the FS represents the imprint of renormalized Fe-d heavy quasiparticles on the H-s orbital via $s$-$d$ hybridization. This phenomenon mirrors the physics of high-$T_c$ cuprates (Zhang-Rice singlets), where correlations drive low-energy weight onto the ligand orbitals \cite{Zhang1988}. 

\begin{figure}[h] 
    \centering
    \includegraphics[width=0.6\textwidth]{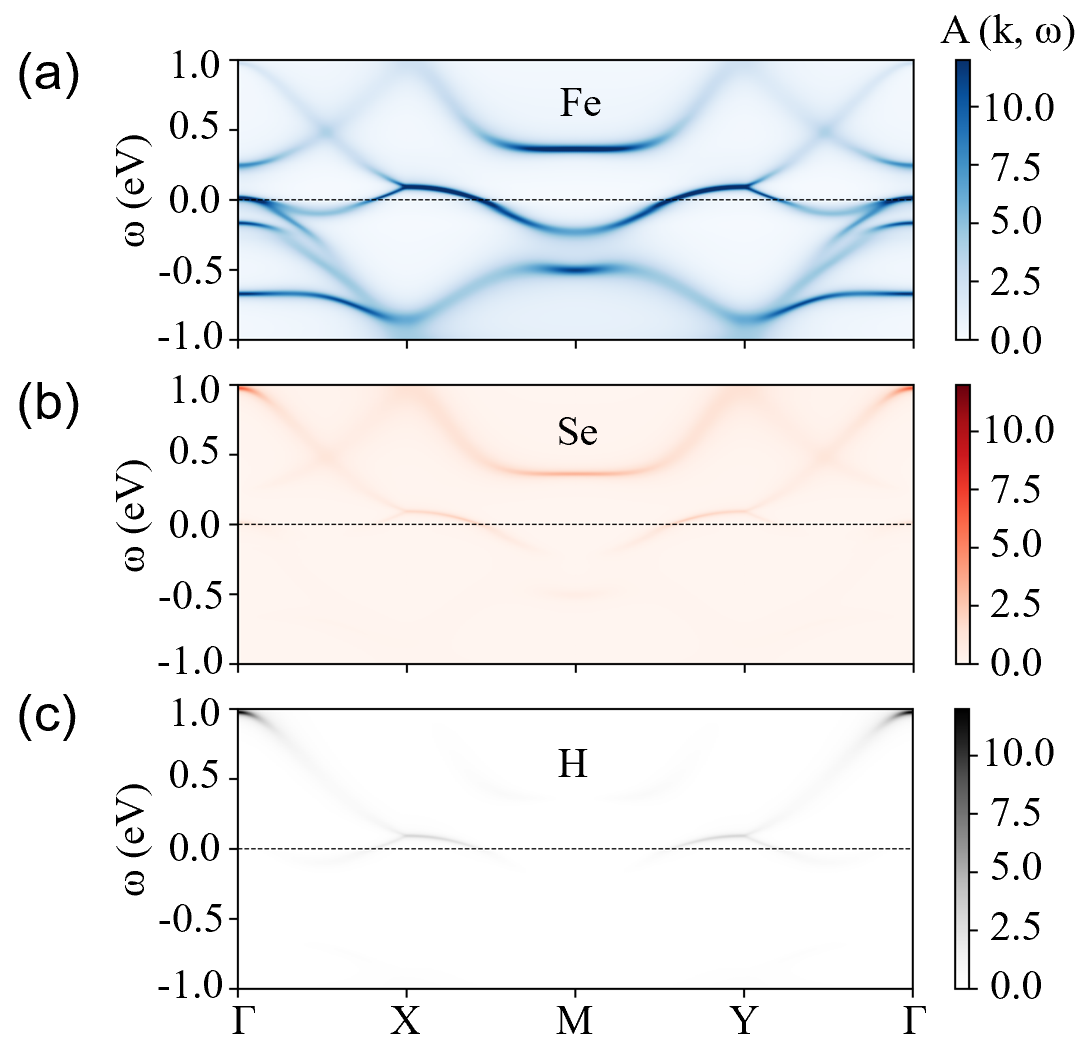} 
    \caption{Element-resolved DMFT spectral function for the band structure of FeSeH projected onto (a) Fe, (b) Se, (c) H.}
\end{figure}

\subsection{Magnetic ground state of FeSeH}
Here, we briefly discuss the magnetism of FeSeH. Magnetism would play a crucial role in the behavior of iron-based superconductors, whereas the interplay between magnetic order and superconductivity being key for understanding their properties \cite{Wang2016,Dai2012,Dai2015,Lumsden2010}. To determine the magnetic ground state of FeSeH, we calculate several typical magnetic orders in iron chalcogenides, including checkerboard, single-striped, and double-striped antiferromagnetic (AFM) configurations, and compare their energies in FIG. 3(c). Among them, the checkerboard configuration has the lowest energy and thus most likely emerges as the experimentally observed magnetic order, consistent with the findings in monolayer FeSe \cite{Wang2016}. Regarding the magnetic anisotropy, the energy difference between the in-plane and out-of-plane orientation for the checkerboard and single-striped configurations is very small, less than 0.1 meV. Whereas for the double-striped configuration, the difference is larger, around 3.6 meV. This indicates that the magnetic anisotropy in FeSeH is relatively weak. The coupling of magnetic order in FeSeH with superconductivity and topology could lead to novel physical phenomena and potential applications, such as the realization of Majorana fermions, which warrant further experimental investigation.

\section{Discussion}
It is worth noting that Table I shows a decrease in both the deformation potential and the EPC strength of the 18th phonon branch after including correlation effects through DMFT. This indicates that the correlation-enhanced EPC is highly mode selective rather than a uniform enhancement across all phonons. In addition, our calculation incorporates DMFT corrections to the EPC only at the $\Gamma$ point, yielding a superconducting $T_c$ of 43.7 K, compared with the 3.6 K obtained from regular DFT. Therefore, regardless whether the DMFT-induced modifications to the EPC at other non-$\Gamma$ $\mathbf{k}$ points lead to an increase or decrease in their individual coupling strength, the overall superconducting $T_c$ of FeSeH remains above 40 K.

The present results can be placed in a broader experimental context. Consistent with the close correspondence between our calculated $T_c$ and that reported for protonated bulk FeSe \cite{Cui2019,Cui2018}, hydrogen incorporation appears capable of driving FeSe into a substantially enhanced superconducting regime. A similar tendency has recently been observed in monolayer FeSe/STO, where hydrogen exposure was found to enhance superconductivity \cite{Xue2024}. While that experiment attributed the enhancement mainly to hydrogen-assisted atomic diffusion and the resulting improvement of interfacial stoichiometry and defect distribution, our calculations provide a microscopic perspective on the superconductivity enhancement induced by hydrogen in FeSe-based systems.
 
In conclusion, our work reveals how hydrogen can actively enhance superconductivity in strongly correlated FeSe-based unconventional superconductors. Using FeSeH as a concrete platform, we show that hydrogen incorporation enhances superconductivity by modifying the electronic structure of FeSe and by introducing high-frequency phonon modes that become strongly coupled to low-energy quasiparticles when electronic correlations are properly included. By combining DMFT-corrected EPC strength with the Migdal-Eliashberg formalism, we find a superconducting $T_c$ exceeding 40 K, substantially higher than that predicted by standard DFT calculations. The fully anisotropic Eliashberg theory further predicts two superconducting gaps in FeSeH, consistent with experimental observations for FeSe. Our results show that many-body correlation-driven orbital renormalization facilitates the participation of hydrogen-derived phonons in electron scattering near the Fermi surface, providing a microscopic route by which hydrogen can promote superconductivity in correlated unconventional superconductors. Our findings provide new insights into the superconducting mechanism of unconventional superconductors. Beyond superconductivity, our calculations suggest that FeSeH favors a checkerboard antiferromagnetic order. The complex interplay among  superconductivity, magnetic ordering, and potential topological properties may offer promising applications in quantum devices and topological quantum computing. Further experimental investigations of FeSeH are highly desirable to test these predictions and to advance the hydrogen-based engineering of FeSe-derived superconductors. Overall, our work constitutes an important step in both the fundamental understanding and practical engineering of unconventional superconductivity.

\section{Methods}
We employ {\it QUANTUM-ESPRESSO} package\cite{Giannozzi2009, Giannozzi2017} to do the DFT calculations. The Brillouin zone is
sampled by $12\times12\times1$ $\Gamma$-centered $\mathbf{k}$ mesh with an energy cutoff of 100 Ry. Norm-conserving pseudopotentials \cite{vanSetten2018} are employed to describe the valence electrons, and the exchange-correlation functionals are treated using the local density approximation (LDA). The DFT results of the superconductivity of FeSeH are obtained using the {\it EPW code} \cite{Ponce2016,Lee2023}. The electron-phonon coupling matrix elements in the Bloch representation are calculated on coarse $12\times12\times1$ $\mathbf{k}$-grid and $4\times4\times1$ $\mathbf{q}$-grid, and further used to interpolate the electron-phonon coupling matrix elements on fine $36\times36\times1$ $\mathbf{k}$-grid and $36\times36\times1$ $\mathbf{q}$-grid using 18 maximally localized Wannier functions that well reproduce bands around the Fermi level. The electron and phonon smearing for calculations of the EPC strength and Eliashberg spectral funciton are set to 0.05 eV and 0.05 meV, respectively. At different temperatures, the anisotropic Eliashberg equations are solved along the imaginary frequency axis up to a cutoff of 0.5 eV with the Coulomb interaction parameter $\mu^{*}$ set to 0.12. The distribution of the superconducting gaps on the Fermi surface are obtained via interpolation on the fine $\mathbf{k}$-grid.

The DMFT-corrected electron–phonon coupling and superconducting $T_c$ calculations follow exactly the same procedure as in Ref. \cite{Du2025}, to which we refer for full methodological details.

\section{Acknowledgments}
 This work is supported by National Natural Science Foundation of China (No. 12025407 and No. 12450401), Chinese Academy of Sciences (No. YSBR-047 and No. XDB33030100) and the Ministry of Science and Technology (No. 2021YFA1400201).
 
\section{Reference}
\bibliography{reference}

\section{Supplemental Materials}

\subsection{I. Other atomic configurations of hydrogenated FeSe}
\begin{figure} [h] 
    \centering
    \includegraphics[width=0.8\textwidth]{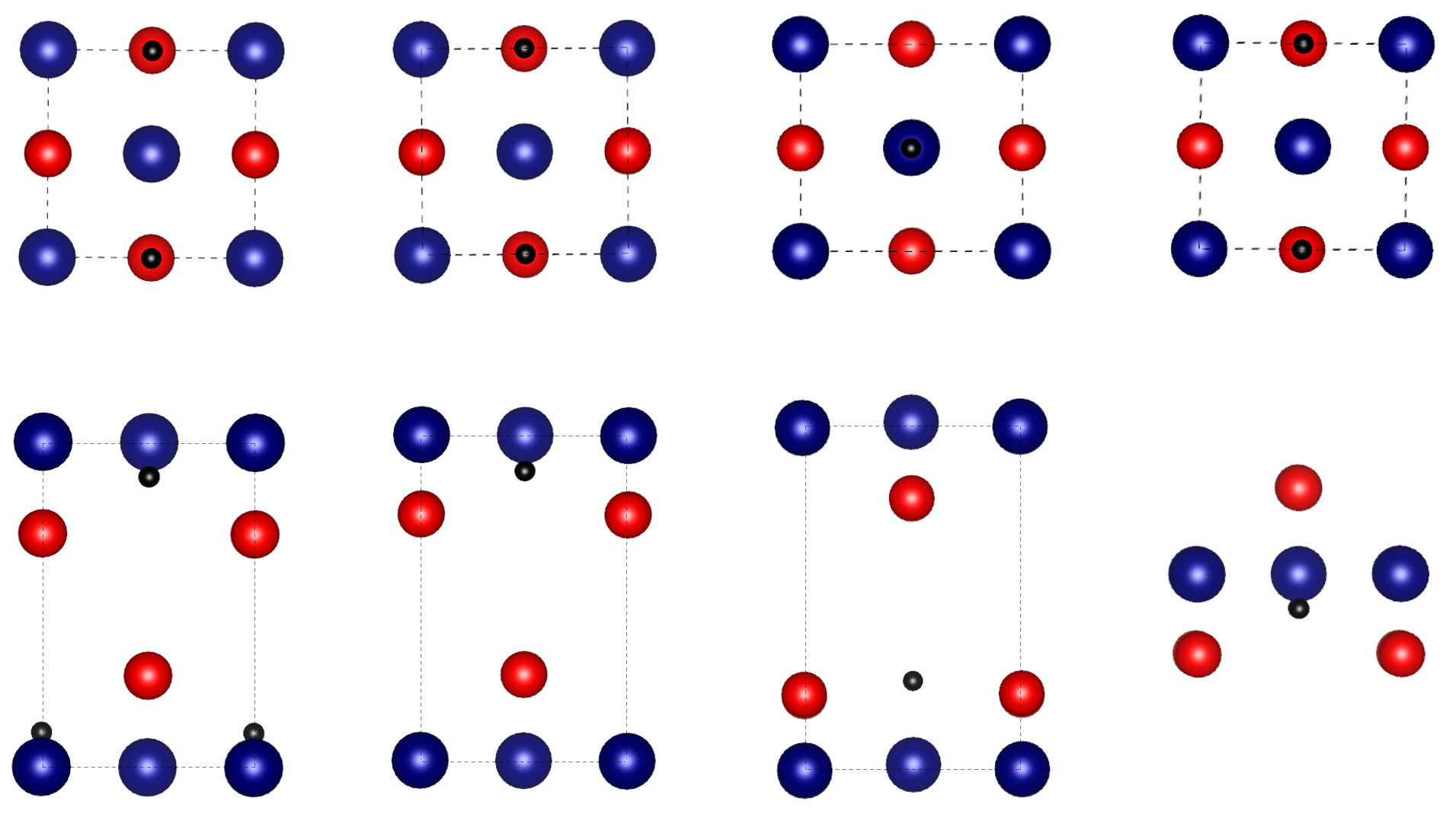} 
    \caption{Other configurations we obtained after structure relaxation (Top: top view; Bottom: side view) but with imaginary phonon frequencies.  The former three are bulk-type, while the last one is monolayer-type.}
    \label{fig:S1}
\end{figure}

\newpage
\subsection{II. Atomic resolved DFT bands of FeSeH}
\begin{figure*} [h] 
    \centering
    \includegraphics[width=0.8\textwidth]{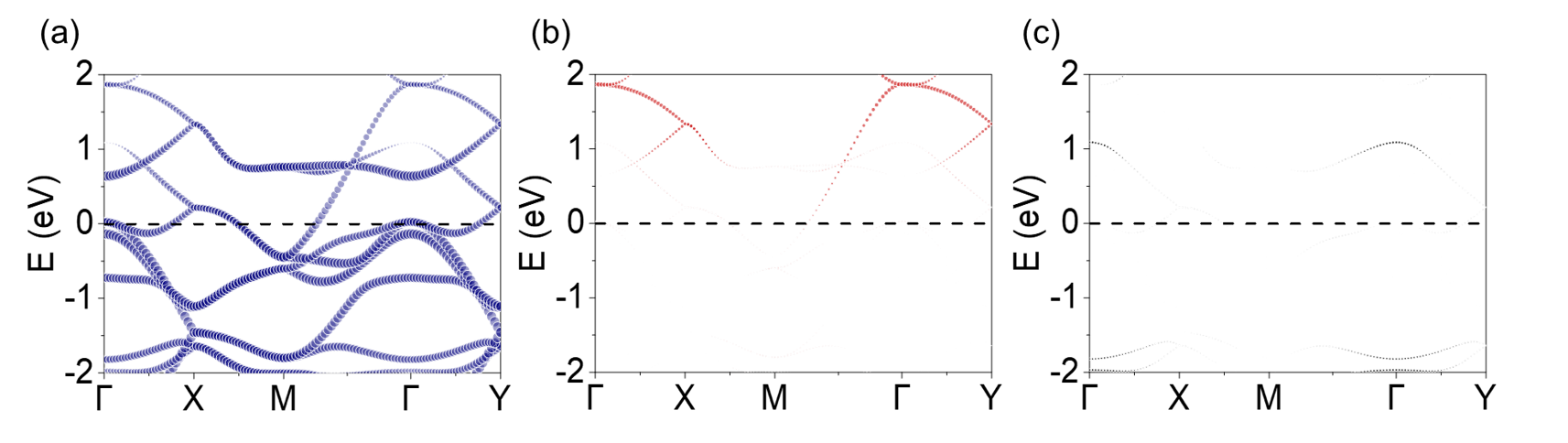} 
    \caption{Atomic resolved bands contributed by (a) Fe, (b) Se, (c) H from DFT calculations.}
\end{figure*}

\newpage
\subsection{III. Changes of spectral functions due to deformation}
\begin{figure} [h] 
    \centering
    \includegraphics[width=0.6\textwidth]{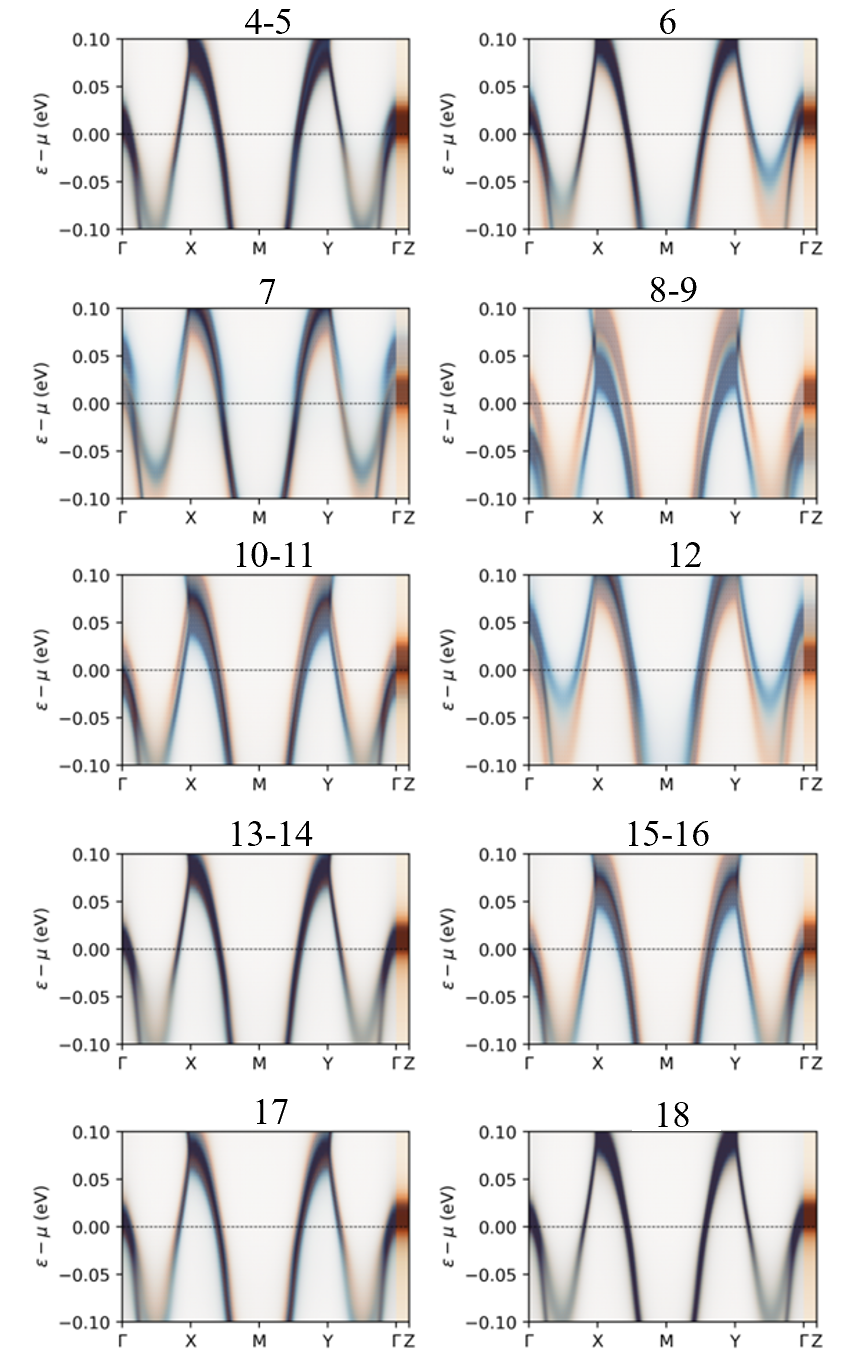} 
    \caption{Changes of DMFT spectral functions near the Fermi surface under the perturbation of the fifteen optical phonon modes. Orange color represents the initial spectral function, while blue represents the one after perturbation. The numbers are indices of the optical phonon modes ranging from 4 to 18.}
\end{figure}

\end{document}